**Teaching Cultural Astronomy to Undergraduates with an Interdisciplinary Frame**
Jarita Holbrook, University of the Western Cape

Cultural Astronomy is interdisciplinary connecting the arts, humanities, social & physical sciences. Data collection methods and theories are used from many disciplines and meld with methods and theories within cultural astronomy. The burden on the student is that to do cultural astronomy research it is necessary to be widely read within and across disciplines. I developed a series of courses that divided cultural astronomy content into broad regions such as Africa, North America, and the Pacific. The courses were structured to accommodate students from all parts of the university, but had to have enough mathematics and science to serve as a general science requirement. The majority of the grade for the course lay with the final project, which included a presentation and a written document. Thus, the course was designed to give the students a foundation for doing this final project that had to be original research. The students rarely opted to collect their own data, instead they re-analysed existing materials. The students learned critical thinking, formulating hypotheses, and how to test their hypotheses, as well as how to understand cultural astronomy data.

**Introduction**
Archaeoastronomy, ethnoastronomy, history of astronomy and cultural astronomy are fascinating fields that have struggled to establish themselves within the academic context. However, there have been successes with the establishment of postgraduate programs over the last two decades such as those established at James Cook University in Australia now in Thailand, Bathspa University in the UK now in Lampeter UK, and Ilia State University in the Republic of Georgia (Orchiston et al. 2011; Nick Campion 2008; Nicholas Campion and Malville 2011; Andrews 2011). Each of these programs are not associated with universities so much as to passionate individuals that are the visionaries of the programs i.e. Nick Campion, Wayne Orchiston, and Irakli Simonia. If they move, the programs move with them. Also, oftentimes the programs are not referred to by their name or university but to these individuals. Having a visionary leader does not mean that they are the only faculty, in fact they have brought together experts from around the world to teach modules and lecture in their areas of expertise. For example at the Sophia Centre in Lampter UK, archaeoastronomers Kim Malville (J. M. K. Malville, Eddy, and Ambruster 1991; J. M. Malville et al. 1998; J. M. K. Malville et al. 2007) and Fabio Silva (Silva and Campion 2015; Silva 2014) are instructors. Perhaps the newest program is at the University of Oklahoma in the USA run by Steve Gullberg and Andrew Munro ("Archaeoastronomy Graduate Certificate" n.d.). As with the other programs mentioned, it is a postgraduate program.

While a faculty member at the University of Arizona, I helped design a cultural astronomy program for undergraduates and postgraduate students. At the postgraduate level, three classes are required to get a graduate minor in a subject. A cultural astronomy class focused on the anthropology of astronomy taught by me and an archaeoastronomy class taught by Dennis Doxtater were two of the three, a third untaught class was to focus on history of astronomy and historical astronomy. The University of Arizona has three times as many undergraduates than postgraduate students (University Analytics and Institutional Research, 2018), thus it is primarily an undergraduate institution. An undergraduate cultural astronomy focused program was possible given the structure of USA universities: to obtain a bachelors degree it is necessary to complete classes in general education as well as classes specific to a student's academic major. The general education classes are offered by all academic departments and are meant to attract majors and non-majors. The cultural astronomy classes were part of the astronomy department and a mix of mathematics, science, and writing to meet the requirement for being an upper level general education class. Thus, these interdisciplinary cultural astronomy classes could be used to meet the science requirement for any academic major at the University of Arizona and when taught, the classes were full.

**What to teach?**
An obvious starting point is an introduction to all the terms used as well as the terms used in adjacent fields such as astrophysics, anthropology, and archaeology. An overview of the history of cultural astronomy going back to the turn of the previous century (S. N. Lockyer 1893; N. Lockyer 1905; J. N. Lockyer and Penrose 1902) highlighting the organizing of societies, conferences, and journals since the 1980s. The most difficult parts of cultural astronomy are covered first - namely the astronomy content. The class was structured socially in that the students divided themselves into celestial groups. Each celestial group had to focus on a celestial body or bodies, and report on their findings as part of weekly discussions. Possible groups included the Sun, the Pleiades, Eclipses, the Zodiac and many more celestial bodies that would have stories or myths attached to them. Each celestial group was required to give detailed scientific explanations of their celestial focus, give visibility updates, and present relevant myths and legends broken down into five-minute reports delivered each week.

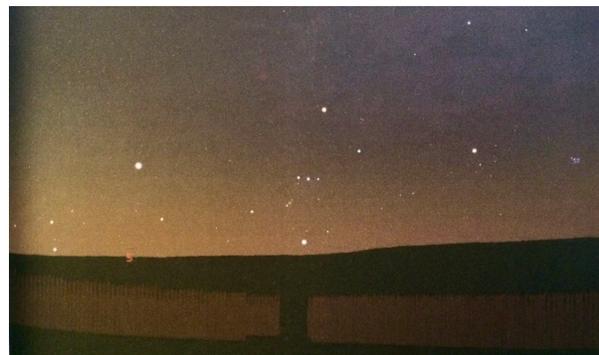

Figure 1: Stellarium image of the night sky with an architectural structure on the ground. Reproduced with permission from Georg Zotti.

Graded items included quizzes, homework assignments, and at the end of the class a final paper and oral presentation. The final paper and presentation were also group efforts of two to four students. In scientific terms, the final paper had to address an original question/hypotheses with at least three supporting arguments to support their conclusions. Two required activities to facilitate the final paper were to pass the USA national ethics examination for doing research on people ("CITI Program – Collaborative Institutional Training Initiative") as well as learning referencing software to reduce plagiarizing. The software available to the students at the time for referencing was Refworks ("ProQuest RefWorks"), now Zotero is more popular ("Zotero | Your Personal Research Assistant").

**Astronomy, Maps, and Mapping**
The first quiz focus was naming the countries on a geographical map of the region under study. The geographical focus of the class changed each year rotating through Africa, the Pacific, and the Americas. It could be envisioned to do similar classes on Asia and another on Europe. From a geographical map, to maps of the night sky, students were tasked with becoming familiar with the night sky with the goal of having a night quiz at a dark location where they were required to identify ten objects correctly. Though planispheres were available, Stellarium (Chéreau 2003) was the best way for students to practice short of going outside at night. Stellarium had the advantage in that the horizon, zenith, and cardinal points are clearly marked and comprehensible. In contrast, a 2D sky map is meant to be understood as if you are lying on the ground looking up thus east is to the left if north is the top, which can be confusing for students. The Stellarium images were easy to understand because the cardinal directions were marked as well as the ground and horizon. For example see Figure 1, a Stellarium image showing an archaeological site on the ground and how the sky looked at that time (Zotti and Neubauer 2016). Other astronomy content includes understanding the phases of the moons, calculating lunar and solar calendars versus the tropical year, the ecliptic, and horizon astronomy.

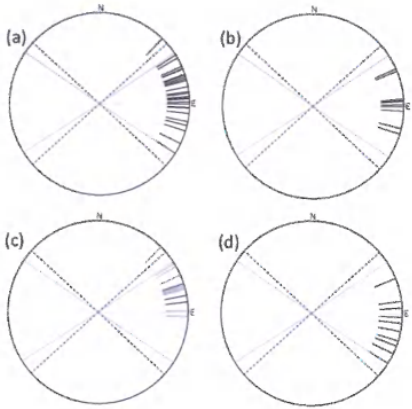

**Figure 2: Example of a Radial Map. The dashed lines show the physical limits of the sun and moon relative to the alignment axes of the churches – the short solid lines. Reproduced by permission from A. César Gonzalez-Garcia.**

In addition to geographical maps of the region of interest and maps of the night sky, students must understand how archaeoastronomers map sites and potential celestial alignments of a site. Archaeology site maps had to be understood in terms of the difference between true north and magnetic north. Two types of alignment maps needed to be understood – the radial maps and the histogram waveform type maps. Figure 2 shows a radial map of the alignments of four different types of churches in Spain reproduced from Gonzalez-Garcia (Gonzalez-Garcia, Belmonte, and Ferrer 2016). The second example,

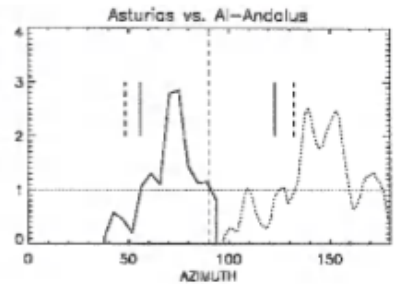

**Figure 3: Example of a histogram map for churches in Spain. The waveforms capture both the direction of alignment and the number of churches with their axes so aligned. The short solid and dashed lines are the physical limits of the sun and moon. Reproduced by permission from A. César Gonzalez-Garcia.**

Figure 3 shows an example of a histogram waveform map for a subset of churches in Spain (Gonzalez-Garcia, Belmonte, and Ferrer 2016).

**Data Collection Methods and Case Studies**
The introductory text used was Fabian's "Patterns in the Sky: An Introduction to Ethnoastronomy" (Fabian 2001). General texts recommended by other cultural astronomers include "The Power of Stars: How Celestial Observations have Shaped Civilization (Penprase 2011)", "History and Practice of Ancient Astronomy (Evans 1998)", and "Exploring Ancient Skies: A Survey of Ancient and Cultural Astronomy (Kelley and Milone 2011). Further data collection methods were covered in relation to case studies (J. C. Holbrook and Baleisis 2008).

Shorter case studies focused were covered in articles, but after several weeks, the next phase was to dive into more complete case studies on specific ethnic groups or the works of particular researchers (e.g. Ammarell 1999; LaPin and Speed 1984). Included were guest lectures tapping into the expertise held within the Tucson community including astrophysicists such as Chris Impey (Impey and Petry 2004), Maghreb astronomy expert Danielle Adams (Adams 2015), and Sirius expert Jay Holberg (Holberg 2007). The enthusiasm brought by these guests breathed life into their topics and showed that cultural astronomy is a vibrant community albeit a small one. In some years, films were included which tended to present data more clearly than articles.

**Student Responses**
The classes were well received by the students. Placing the mathematics and astronomy topics at the beginning of the class had the effect of driving away the students that thought it would be an easy class. Being critical and thinking about what's missing in each work presented was part of the training. The lessons were discussion based allowing for thorough exploration. The students came to understand that one site with alignments was not enough to prove a relationship between people and the sky, and that other sites with the same alignments along with other types of data such as related artifacts or ethnographic support were needed to solidify proof. That alignments had to take into account horizon features as well as the topography of the site, was another important lesson. Students did not like the circular maps because the authors didn't always indicate from which point they were doing the measuring. The night sky activity and quiz with laser pointer was exciting for the students.

In terms of their final project, most students chose to bring together existing data to answer their question rather than collect their own data. Perhaps this is not unexpected given that the area of interest was relatively distant such as Africa or the Pacific, however, Tucson has an African refugee population as well as students from those locations that could have been surveyed on their sky knowledge as part of an original project.

**Conclusions**
The interdisciplinary undergraduate classes introduced students to cultural astronomy, the tools of cultural astronomy, and gave them the opportunity to do a small cultural astronomy research project. For SEAC and the international cultural astronomy community, teaching undergraduates may or may not lead to future researchers; however, it fosters the possibility of younger researchers becoming active, which can lead to new developments if they spend more of their academic life undertaking cultural astronomy research. The structure of the university in the USA feeds a demand for cultural astronomy classes at the undergraduate level. This may not be the case in other countries where there aren't general education requirements. Finally, cultural astronomy is an attractive field for students regardless of the nearly nonexistent job prospects.